\documentclass[sigconf]{acmart}

\AtBeginDocument{%
  \providecommand\BibTeX{{%
    \normalfont B\kern-0.5em{\scshape i\kern-0.25em b}\kern-0.8em\TeX}}}

\copyrightyear{2024}
\acmYear{2024}
\setcopyright{rightsretained}
\acmConference[CHI '24]{Proceedings of the CHI Conference on Human Factors in Computing Systems}{May 11--16, 2024}{Honolulu, HI, USA}
\acmBooktitle{Proceedings of the CHI Conference on Human Factors in Computing Systems (CHI '24), May 11--16, 2024, Honolulu, HI, USA}
\acmDOI{10.1145/3613904.3642717}
\acmISBN{979-8-4007-0330-0/24/05}

\begin{document}

\title[From Prisons to Programming]{From Prisons to Programming: Fostering Self-Efficacy via Virtual Web Design Curricula in Prisons and Jails}


\author{Martin Nisser}
\affiliation{%
  \institution{MIT CSAIL}
  \streetaddress{32 Vassar St.}
  \city{Cambridge}
  \state{MA}
  \country{USA}}
\email{nisser@mit.edu}

\author{Marisa Gaetz}
\affiliation{%
  \institution{MIT}
  \streetaddress{32 Vassar St.}
  \city{Cambridge}
  \state{MA}
  \country{USA}}
\email{mgaetz@mit.edu}

\author{Andrew Fishberg}
\affiliation{%
  \institution{MIT}
  \streetaddress{32 Vassar St.}
  \city{Cambridge}
  \state{MA}
  \country{USA}}
\email{fishberg@mit.edu}

\author{Raechel Soicher}
\affiliation{%
  \institution{MIT}
  \streetaddress{32 Vassar St.}
  \city{Cambridge}
  \state{MA}
  \country{USA}}
\email{raechel@mit.edu}

\author{Faraz Faruqi}
\affiliation{%
  \institution{MIT CSAIL}
  \streetaddress{32 Vassar St.}
  \city{Cambridge}
  \state{MA}
  \country{USA}}
\email{ffaruqi@mit.edu}

\author{Joshua Long}
\affiliation{%
  \institution{University of Massachusetts, Lowell}
  \streetaddress{220 Pawtucket St.}
  \city{Lowell}
  \state{MA}
  \country{USA}}
\email{joshua_long@uml.edu}


\renewcommand{\shortauthors}{Nisser, Gaetz, Fishberg, Soicher, Faruqi, Long}

\begin{abstract}

Self-efficacy and digital literacy are key predictors to incarcerated people’s success in the modern workplace. While digitization in correctional facilities is expanding, few templates exist for how to design computing curricula that foster self-efficacy and digital literacy in carceral environments. As a result, formerly incarcerated people face increasing social and professional exclusion post-release. We report on a 12-week college-accredited web design class, taught virtually and synchronously, across 5 correctional facilities across the United States. The program brought together men and women from gender-segregated facilities into one classroom to learn fundamentals in HTML, CSS and Javascript, and create websites addressing social issues of their choosing. We conducted surveys with participating students, using dichotomous and open-ended questions, and performed thematic and quantitative analyses of their responses that suggest students' increased self-efficacy. Our study discusses key design choices, needs, and recommendations for furthering computing curricula that foster self-efficacy and digital literacy in carceral settings.

\end{abstract}



\begin{teaserfigure}
  \includegraphics[width=\textwidth]{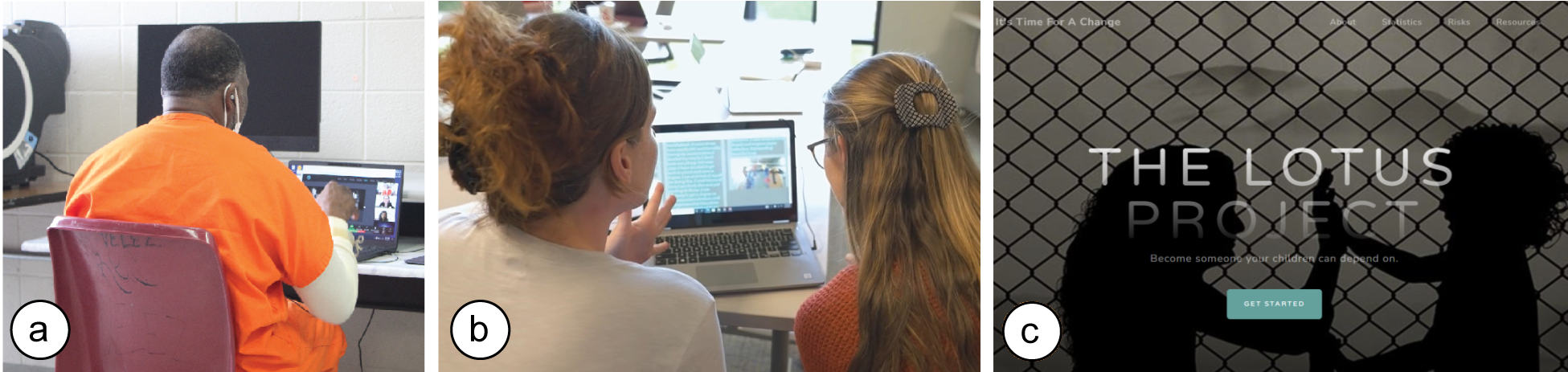}
  \caption{Scenes from our virtual, mixed-gender web design program for incarcerated people. Men and women learn web design fundamentals, using these to create websites addressing a social issue of their choosing to foster digital literacy and self-efficacy. a) A student from a men's correctional facility participates in class via Zoom. b) A student from a women's correctional facility presents her website to an in-person facilitator. c) A website created by a student that addresses addiction to drugs and alcohol. }
  \Description{Two photographs of incarcerated students working at their laptops, and a screenshot of a webpage.}
  \label{fig:teaser}
\end{teaserfigure}

\maketitle

\section{Introduction}

{As of 2021, the United States was reported to have the highest incarceration rate in the world {\cite{widra2021states}}. Furthermore in 2021,} 0.7\% of the U.S. population was in a prison or jail, while over 2\% of the population was under some form of carceral supervision, including probation or parole~\cite{BJS21}. The negative impact of mass incarceration has wide-ranging consequences both economically and socially; due to legal discrimination, societal stigma, and intersecting issues of race, gender and economic class, incarcerated people are one of the most marginalized populations in society~\cite{western2010incarceration}. Among their needs are better paths to digital literacy, education and self-efficacy, topics for which researchers in human-computer interaction (HCI) are well positioned to contribute. However, research addressing incarceration remains largely under-explored in HCI, in part due to difficulties in accessing prisons and jails. In this work, we report on a 2-year study of a custom virtual web design program for incarcerated people, one of the first of its kind, designed to foster digital literacy and self-efficacy to promote success in the workplace post-release. 

Correctional education weighs heavily on issues that are important to the computer science and HCI communities, intersecting on issues in technology, education, and social justice. The racial and ethnic makeup of incarcerated populations are significantly skewed; African American and Hispanic people are incarcerated at 5.6 and 1.8 times the rate of white Americans, respectively \cite{mauer2007uneven}. The loss in earnings associated with imprisonment for an individual has been found to range between 10\% and 30\% \cite{western2002impact}, further exacerbating economic inequality. Incarcerated students participating in post-secondary education are often first-in-family to do so \cite{farley2016engaging}. In addition, parental incarceration has shown to have a detrimental impact for child well-being for the 2.6 million children in the United States with a parent who is incarcerated~\cite{poehlmann2021developmental}. Research and programming that inform how to ameliorate the rates of incarceration are inextricably linked to advancing social equity, child well-being, and economic health. 


A central cause of mass incarceration is recidivism, the rate at which a person released from a correctional facility engages in actions that result in rearrest, reconviction or a return to a facility. In the period 2005-2014, an estimated 68\% of people released from a jail or prison in the U.S. were arrested within 3 years, 79\% within 6 years, and 83\% within 9 years \cite{BJS18}. However, a study aggregating 37 years of research (1980-2017) on correctional education showed that people who participate in post-secondary educational programming while incarcerated are 28\% less likely to return to prison~\cite{bozick2018does}. This study also reports that those participating in any form of correctional education are 12\% more likely to find post-release employment, but this effect was not statistically significant. In practical terms, education does not \textit{guarantee} employment post-release, but it is associated with a significant and substantial reduction in recidivism. 

This observation has prompted researchers to investigate other mediating factors that may cause correctional education to help students find employment opportunities post-release. One such factor to be revealed is the tendency for education in carceral settings to stimulate general \textit{self-efficacy}, which is well-documented as a key predictor in preventing recidivism. Studies show that higher self-efficacy scores are correlated with lower recidivism rates for crimes including driving under the influence (DUI) \cite{wells2000self}, sex-related crimes \cite{pollock1996self} and drug-related crimes \cite{yamamoto2013relationship}. Allred et al.~\cite{allred2013self} also show that academic accomplishments arising from participating in the same college-level course have a larger effect on self-efficacy for incarcerated students than for students who are not incarcerated. Other studies reveal complementary effects, including that incarcerated people scoring higher on certain forms of self-efficacy are more likely than those with lower scores to choose to participate in educational programs~\cite{roth2017academic}.

In addition to general self-efficacy, \textit{digital literacy} (or computer self-efficacy) is key to achieving post-release employment for incarcerated people today~\cite{farley2016engaging}. Inequalities in digital literacy disproportionately affect vulnerable and marginalized populations, including formerly incarcerated persons, who experience compound effects, including low incomes and education rates. Moreover, they typically lack access to computers and the internet while incarcerated, depriving them of the necessary digital skills to navigate the economic, social, cultural, personal, and health-related resources that are now embedded in digital technologies~\cite{reisdorf2022locked}. The unemployment rate for formerly incarcerated people is approximately 27\%, nearly 5 times higher than the general U.S. population, suggesting systemic labor market inequality \cite{couloute2018out}. In addition, joblessness\textemdash a measure that includes anyone who does not have a job, whether they are looking for one or not\textemdash is significantly higher for incarcerated people, further implicating self-efficacy as an important factor to boosting labor market participation. Almost two thirds of formerly incarcerated people becoming employed today enter jobs that are typically available to people with little or no education~\cite{carson2021employment}. These include, but are not limited to waste management, manufacturing, and construction. Technological and digital literacy play a decisive role in job searches today~\cite{dillahunt2020positive}. Despite this, technological limitations and bureaucracy hinder educational opportunities in computing for incarcerated individuals, leading to them struggling to keep up with the requisite skills in the modern labor-market~\cite{heicklen2019coding}.

Self-efficacy and digital literacy have been revealed to play critical roles in preventing recidivism and fostering success in the workplace post-release. However, there are significant challenges to accessing, tracking, and measuring data related to educational programs for incarcerated people. These challenges include IRB approvals, facility approvals, relocation of students between facilities, security issues regarding computers and the internet, involuntary removal of students from courses due to misconduct, and the psychological or emotional strain due to the incarcerated experience that prevents students from completing an educational program. These challenges have significantly limited the ability of researchers to study the effects of educational programs on self-efficacy and digital literacy in carceral settings. Through three years of working with correctional facilities to deliver internet-enabled virtual programming behind the wall, we report on first-hand experience of navigating these concerns to both instruct and study one of the first virtual web programming courses in the United States. 

In this paper, we introduce \textit{Brave Behind Bars}, a novel web design program designed for incarcerated people, and assess its ability to foster self-efficacy and digital literacy as key metrics to ensure successful reentry post-release. We first introduce the program, a college-accredited web design class taught synchronously and remotely, via Zoom, for 12 weeks across 5 correctional facilities in the United States. The first 6 weeks of the curriculum centers on the fundamentals of HTML, CSS and Javascript, and the latter 6 weeks center on a capstone project where students apply taught material to build websites addressing a social issue of their choosing. We follow with a report on two studies, one from 2022 and one from 2023. The 2022 study reports on qualitative feedback from students' experience of the program. Through a thematic analysis, these indicated increases in both students' self-efficacy and computer literacy, as well as engendering positive outlooks with regard to future aspirations. Based on these indicators, a follow-up study was designed for 2023 to corroborate these results quantitatively. Statistical analyses were used to determine whether students' general and computer self-efficacy at the end of the course were improved relative to the start of the course. The aggregated findings demonstrate increased self-efficacy in both categories, but without statistical significance. We discuss the difficulties for statistical analyses to corroborate qualitative feedback, a primary cause of which is the challenge of establishing sufficiently large sample sizes in this area of work~\cite{barros2021overall}. This statistical shortfall is mirrored elsewhere in corrections research \cite{bozick2018does,yamamoto2013relationship} where formidable challenges accessing and approving data collection prevail. Together with our qualitative analysis, which suggests strong positive associations between educational programming and self-efficacy, our quantitative analysis adds important findings that help inform the design of correctional programs to increase self-efficacy as a method to combat recidivism. We end by discussing needs and recommendations for introducing computing curricula in carceral settings in the future.

\section{Related Work}


\subsection{Digital Technology in Correctional facilities}

{
Hawley et al.{~\cite{hawley2013prison}} highlight that a significant proportion of incarcerated people have low education levels, exhibit high drop-out rates, lack lifelong learning competencies, and exhibit low motivation. As Barros et al.{~\cite{barros2023learning}} report, prison learning environments must respect different types of learning and paces, while encouraging participation. However, while incarcerated people are a heterogeneous group{~\cite{czerniawski2016race}}, their diverse learning needs are not typically accounted for in prison environments{~\cite{hopkins2014prisoners}}. 

Analyzing the requirements for internet-enabled learning in prison contexts, previous studies have outlined the need for courses to promote skills development, interactivity, self-confidence, and motivation{~\cite{moreira2017adult}}, echoing findings for adult learners in general{~\cite{monteiro2016inclusive}}. Identifying teaching practices to foster these effects, a study in UK prisons by Gray et al.{~\cite{gray2019transformative}} reports that active classroom participation is key to fostering students' sense of self-determination and confidence. Other work on digitization in prisons suggests a link between digital contact with the outside world and greater self-esteem{~\cite{lindstrom2020smart}}, and the capacity for internet-enabled audiovisual communication to grant access to programming that increases post-release circumstances, facilitating rehabilitation and reducing recidivism{~\cite{mckay2022carceral}}.

Reporting in 2018, Reisdorf and Rikard{~\cite{reisdorf2018digital}} discuss the scarcity of US computer literacy programs inside correctional facilities, and state that none cover how to navigate the internet or use modern technologies. The ubiquitous digitization of society over the past 20 years has led researchers to suggest that denying access to these tools while incarcerated effectively constitutes a second punishment, resulting in incapacitation and further social exclusion post-release{~\cite{zivanai2022digital}}. This concern is compounded following the United Nations declaring internet access a human right{~\cite{kravets2011report}}. Now, governments have begun a shift in policy. In the UK, groups such as Justice Digital, working with the Ministry of Justice, and The Center for Justice, a domestic non-profit, are exploring ways to drive digital change in UK prisons{~\cite{CSJ}}. In the United States, the California Department of Corrections and Rehabilitation announced in 2022 a forthcoming roll-out of 30,000 secure laptops to incarcerated students with access to pre-approved internet sites{~\cite{cdcr}}.

With digitization set to expand in carceral environments, there is high need to study implementations that meet key metrics linked to positive post-release outcomes, such as self-efficacy. However, many studies investigating the need for digitization in correctional facilities are review-based, observational, or theoretical in nature, without directly evaluating the efficacy of new interventional measures inside prisons. Two key reasons for this to date are digital security and federal regulation. First, access to computers and internet in U.S. correctional facilities have been tightly controlled, and second, incarcerated people's classification as a federally-protected population of research participants creates practical hurdles to research in this area{~\cite{research}}. While a number of studies assess the benefits of digitization in general, few studies assess the efficacy of digital-based computer programs in particular. In this paper, we describe a new digital literacy program, conducted virtually in five correctional facilities, and use surveys to suggest what specific course designs may contribute to success in terms of student self-efficacy. 
}

\subsection{Mitigating incarceration in HCI}

HCI researchers have previously cited racial disparities in prisons to argue that "all HCI research must be attuned to issues of race; participation of underrepresented minorities must be sought in all of our activities"~\cite{ogbonnaya2020critical}. Despite this, Verbaan et al.~\cite{verbaan2018potentials} report that prison contexts are under-explored from an HCI perspective; only a handful of publications actually perform work within that context, and the incorporation of technology remains in its infancy in terms of incarcerated people's access. {As discussed, primary reasons for this are digital security and federal regulation.} However, research addressing issues on incarceration, a topic that has become central to discussions around justice in the United States, has in the last few years been increasingly explored in HCI research. 

Ogbonnaya-Ogburu et al.~\cite{ogbonnaya2019towards} found that returning citizens who experienced long prison sentences emerge with almost no digital literacy skills, and piloted a six-week digital literacy course in which they found that returning citizens were hungry for digital literacy skills and differed in their needs in key ways from other marginalized groups. Other researchers examined participatory design processes of a virtual reality (VR) reentry training program with a women’s prison to create narratives that reflect the group's particular challenges~\cite{teng2019participatory}. However, the scope of this work did not include an evaluation of its efficacy. Anuyah et al.~\cite{anuyah2023characterizing} introduced a framework based on Maslow's hierarchy of needs to conceptualize how lack of digital literacy and technology access impose barriers on incarcerated and other marginalized peoples. Interviews conducted with 75 women transitioning out of prison showed how low self-efficacy and digital literacy are major issues faced by women-in-transition in particular~\cite{seo2021informal}. More generally, digital literacy has been identified as a key barrier to accessing social services for formerly incarcerated people and other vulnerable community members~\cite{anuyah2023exploring}. In our work, we report on a web programming course that we designed and taught in 5 correctional facilities, and whose outcomes on self-efficacy and digital literacy we have analyzed, providing a vital link between incarcerated people's experiences and HCI practitioners.

Other HCI researchers have addressed incarceration from a different approach. While our work addresses interventions that may help people stay out of prisons, other researchers have worked on mitigating bias in sending people into them. HCI researchers have  explored the perceptions of algorithmic unfairness in traditionally marginalized communities, for whom consequences can include arrest~\cite{woodruff2018qualitative}. \textit{Street-level Algorithms}~\cite{alkhatib2019street} investigated judicial bail–recommendation systems, interrogating biases that included disproportionately recommending jail for people of color. Veale et al.~\cite{veale2018fairness} interviewed machine learning pracitioners to gauge how designers can imbue values like fairness and accountability into algorithmically-informed public decisions on issues including sentencing and resource allocation within prisons. Bauman et al.~\cite{bauman2018reducing} developed machine learning-driven intervention models to connect social and mental health workers with at-risk populations in need of care, in order to avoid incarceration. Rather than mitigating algorithmic bias in arrests and sentencing, our contribution complements the existing body of HCI work by revealing interventions that foster successful reentry once sentences are served.

\subsection{The Case for Self-Efficacy and Digital Literacy}

Prison-based education programs are effective for improving reentry outcomes. In a 2013 meta-analysis, researchers found that incarcerated adults who participated in reading and/or mathematics courses had a 13\% lower rate of recidivism compared to those who did not~\cite{RR-266-BJA}. A more recent meta-analysis spanning research published from 1980-2017, found an even more significant result: incarcerated students in correctional education programs were 32\% less likely to recidivate~\cite{bozick2018does}. In this same analysis, researchers found that incarcerated people who participated in correctional education had 12\% higher odds of obtaining post release employment compared to those who did not. However, it is unclear whether the association between prison-based education programs and these positive outcomes is a causal one. 

It may be the case that the positive link between correctional education and reentry outcomes is driven by psychosocial characteristics or cognitive skills, as is suggested by research in traditional post-secondary academic settings. Self-efficacy, or the belief in one’s ability to execute behaviors necessary to meet a goal~\cite{bandura_social_1986}, and digital literacy, the competencies needed to perform tasks in digital environments, are two possible mediating factors that have been associated with improved academic performance. A systematic review synthesizing research from 2003 to 2015 showed that self-efficacy was moderately positively correlated with academic performance~\cite{honicke2016influence} and this relationship extends to asynchronous, online courses~\cite{yokoyama2019academic}. Similar effects have been reported for digital literacy and performance~\cite{yustika2020digital}, as well as for digital literacy and students' intentions to pursue further education~\cite{mohammadyari2015understanding}. Overall, the benefit of high self-efficacy and digital literacy skills for academic achievement is well-supported by research in non-carceral academic settings.

A limited number of studies have investigated psychosocial characteristics or cognitive skills of incarcerated populations due to the traditional emphasis on outcomes like recidivism. Those studies have found a positive relationship between postsecondary correctional education and self-efficacy, cognitive skills, self-evaluations, and self-esteem. More focused research indicates that self-efficacy is a key mediator of the effect of correctional education on both proximal outcomes like academic success and more distal outcomes like recidivism~\cite{allred2013self,jones2019literacy}. Reentry programs with a focus on digital literacy acquisition have led to significant reductions in recidivism~\cite{withers2015corrections}. However, with research aiming to establish relationships between education and outcomes, a key factor that distinguishes studies with incarcerated people is the challenge of setting up research programs and acquiring large sample sizes. To name one example, a recent meta-analysis~\cite{bozick2018does} found no significant effect of computer-assisted instruction on academic performance (compared to traditional instruction), however the dataset was too limited to draw conclusions about this relationship. Our study is an important contribution to the sparse literature investigating both psychosocial factors and digital literacy skills in the context of incarceration.





\section{The web design program}

In this section, we introduce the structure of our web design program, \textit{Brave Behind Bars}, and detail the challenges involved with studying educational programs in correctional environments. Our program is a college-accredited web design course designed for incarcerated people. Intended to foster digital literacy and self-efficacy, students learn to build websites that address a social issue of their own choosing. To humanize the educational experience, men and women from separate, gender-segregated facilities join one virtual classroom to learn and work together. The program is taught synchronously and remotely, via Zoom, in 5 correctional facilities across the Eastern United States. The program is accredited for students through three partnering universities: Georgetown University, Benjamin Franklin Institute of Technology, and Washington County Community College. 


\subsection{Structure and curriculum}

The program runs for 12 weeks, meeting twice weekly for class, and once weekly for office hours. Each class is 2 hours long, consisting of a 30 minute lecture followed by 90 minutes spent in breakout rooms with a Teaching Assistant (TA). Each breakout room was staffed by 1 Teaching Assistant, working with 2-4 students to implement and debug code. TAs were recruited predominantly from our institution's computer science department, and included PhD students, faculty, staff, and alumni. We also invited formerly incarcerated graduates of our program to TA, whose input helped align our program's teaching to the current students' needs. The choice of using breakout rooms, combining low student-teacher ratios with project-based coursework, is modelled on studio-based instruction. Within the HCI community, researchers have studied how to optimally teach computer science, with studies demonstrating that students overwhelmingly endorse studio-based design to learning involving project-based open-ended design problems \cite{docherty2001innovative}. The studio course's attendant teaching mode of mentor and mentee can be found wherever design is taught \cite{docherty2001innovative}. For courses on subjects like web programming, it is this design orientation that makes HCI-related courses exceedingly suitable for Studio-based instruction \cite{hundhausen2011prototype}. To foster Studio-like mentor-mentee relationships, we combined our project-based curriculum with dedicated recruitment drives to achieve low student-teacher ratios. Our 2022 class had 16 Teaching Assistants and 45 students enrolled at course start, for an average breakout room student-teacher ratio of 2.8:1. In 2023, the program had 21 Teaching Assistants, with 57 students enrolled at course start, for an average breakout room student-teacher ratio of 2.7:1. In correctional facilities in particular, studies reveal the importance of "meeting the students where they were" in terms of academic abilities, and of providing detailed and constructive feedback~\cite{link2016breaking}. Our program's template for recruiting large numbers of outside Teaching Assistants may also mediate broader benefits; as Link et al.~\cite{link2016breaking} report, one way to diminish the ``us versus them" attitudes exhibited by non-incarcerated people is to allow these individuals to engage in shared learning experiences that create a sense of community. In summary, our student-teacher ratios were designed to facilitate individualized teaching that catered to the breadth of student backgrounds and fostered an inclusive learning environment. The curriculum and individualized teaching was designed to support learners with no previous computer experience and required no academic prerequisites. 

\begin{figure*}[t]
  \centering
\includegraphics[width=0.98\textwidth]{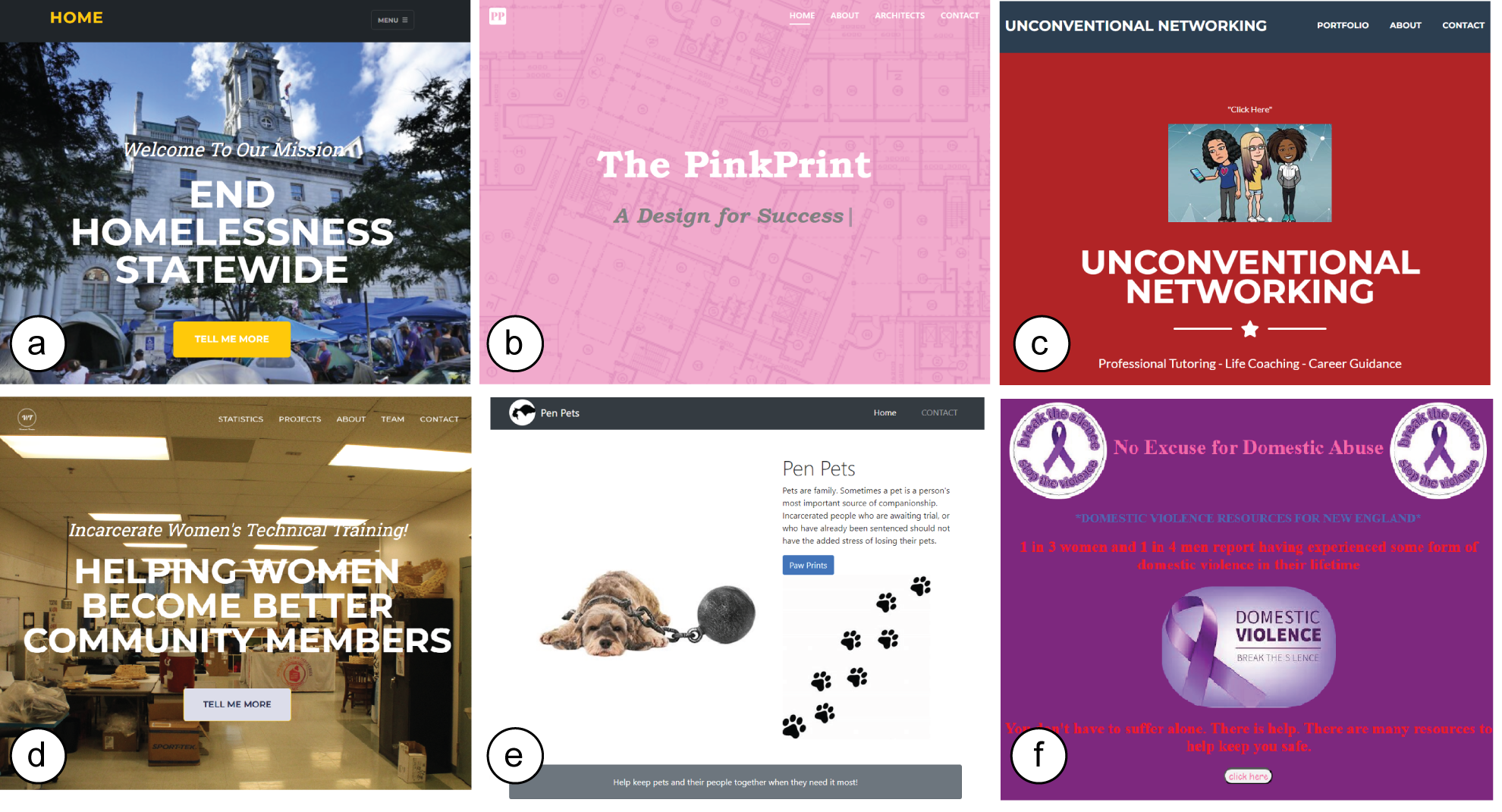}
  \caption{Homepages from six websites created by incarcerated students in our 2023 study that provide resources and tools for use by themselves and the communities to which they will return. a) \textit{End Homelessness Statewide} provides resources for unhoused people to find temporary and permanent shelter. b) \textit{The PinkPrint} provides a "blueprint" that addresses the state of invisibility of justice-involved women by providing educational and gender-responsive resources that incarcerated women can turn to. c) \textit{Unconventional Networking} provides a platform allowing professionals to mentor and guide at-risk youths in their careers. d) \textit{Helping Women Become Better Community Members} provides resources to help women navigate gender and pay disparities when entering technical trades. e) \textit{Pen Pets} helps people entering prisons and jails find homes for their pets while their sentence is served. f) \textit{No Excuse for Domestic Abuse} outlines current statistics regarding domestic violence and provides resources for victims to seek help.}
  \Description{Screenshots of six student websites.}
  \label{fig:websites}
\end{figure*}

The first 6 weeks of the 12-week curriculum centers on the fundamentals of HTML, CSS and Javascript, including branching statements, loops, and functions. The latter 6 weeks
center on a capstone project. For this, students apply their learnings to build websites addressing a pressing social issue of their own choosing to affect impact in their communities. Popular issues chosen by students included domestic violence, gun control, and addiction\textemdash issues which many students have first-hand experiences with. Their websites centered around offering support, outlining factual information, and raising awareness about their topic, using their technical skills to provide a vital service while reconnecting with the communities to which they will eventually return. The homepages from six websites programmed by students in our 2023 cohort are shown in Figure~\ref{fig:websites}. As a college-accredited program, all passing students can utilize their credits to enroll in one of the degree-granting partner universities post-release.

\subsection{Planning and implementation}

To foster transparency and encourage further research in correctional education, we report on some of the logistical, security, and administrative challenges with bringing educational programming to facilities, and with collecting data on these programs. These challenges include enabling computer instruction, internet access, and bringing men and women into a virtual room to learn together. While data for these topics is sparse, and the landscape is rapidly changing, we summarize key challenges and place them in context. 

\subsubsection{Approving a class in correctional facilities}
\hfill \break
One primary challenge to deploying computing curricula in carceral settings is the lack of access to computers and internet in U.S.~correctional facilities. While some correctional facilities in Australia~\cite{farleyhopkins}, Belgium~\cite{gilna2017prisoncloud}, and Finland~\cite{Finland} have begun to allow internet access, this access remains extremely limited in U.S.~correctional facilities~\cite{reisdorf2022locked}. A 2011 survey reported that only 7\% of postsecondary correctional education programs were delivered using internet, while respondents from all 43 states reported delivering on-site instruction~\cite{gorgol2011unlocking}. If limited internet access is available, it is often not accessible to students through computers. Instead, access is provided through sanctioned tablets, often sourced from for-profit providers such as JPay \cite{finkel2019more}, and may consequently be expensive to use~\cite{law2018companies}. While these tablets can provide educational content, they are impractical for teaching the skills our program aims to deliver, including the use of a mouse and keyboard, navigating common operating systems, and performing programming-based coursework. Our program was carried out in correctional facilities with existing computer laboratories. However, to enable our program, one facility provided internet access to its residents for the first time in its history, with plans to continue doing so in the future. 

Another logistical necessity for providing instruction in correctional facilities involves training and security clearances for teachers; this applies to virtual programs too. Our program's instructors and teaching assistants must complete training programs on the Prison Rape Elimination Act (PREA) and other topics, up to 8 hours in length, on a per-facility basis. Teachers may also be required to undergo individual background checks administered by participating facilities. When formerly or currently incarcerated people wish to participate as teaching assistants, facilities may have specific instructions. These include restricting their access to students from their current or former facility, or requiring supervision by a course instructor.

In addition to challenges facing course staff, there are additional obstacles to approving a mixed-gender course. Data on this topic is sparse, however to our knowledge, our program is one of the first mixed-gender courses in correctional facilities in the United States. Historically, since correctional facilities are predominantly gender-segregated, and virtual programming remains uncommon~\cite{gorgol2011unlocking}, mixed-gender courses are nearly unprecedented. With the normalization of virtual instruction in traditional academic settings, its adoption in carceral settings may grow, however their historical precedence of gender-segregation may pose a challenge to mixed-gender classrooms even in a virtual capacity. Facilities that do permit virtual mixed-gender courses often impose extra security protocols, such as requiring the meeting chat to be disabled and for students to not be left unattended in breakout rooms without supervision.


\subsubsection{Approving a study in correctional facilities}
\hfill \break
Following the approval of an educational curriculum in a correctional setting, securing approvals for research studies presents another challenge. Incarcerated people are a federally-protected population of research participants \cite{research}. IRB approvals are therefore required for any study involving incarcerated people, including program evaluation studies like ours. In addition to IRB approvals, all of the participating correctional facilities also need to approve the studies. Once a study like ours has begun, the carceral environment can cause a variety of disruptions for students that impact the size and quality of the dataset. These include transfer of students out of a facility, removal of a student by the facility due to misconduct, or student absences due to the psychological or emotional strain of incarceration. For example, of the 45 students enrolled in the 2022 program, 7 (15.6\%) were unable to complete the program due to facility transfer or other involuntary reasons, including other mandatory programming, medical reasons, or work conflicts. An additional 4 students (8.9\%) left the course due to unknown reasons. Similarly, in 2023, 9 students (15.8\%) of the 57 enrolled students left the course due to facility transfer or other involuntary reasons, and an additional 6 (10.5\%) left the course for unknown reasons before the conclusion of the study. While these figures report on the causes of why students may involuntarily discontinue class, these same causes can also lead to temporary student absences that disrupt their educational experience despite remaining enrolled.


\section{Methodology}

{In 2022, we designed a post-course survey to asses the students' experiences of the program, focusing on self-efficacy. A thematic analysis of the survey, discussed in the results, suggested strong increases in self-efficacy, both in general and with regard to digital literacy specifically. In response, we isolated general and computer self-efficacy as measures for a quantitative follow-up study in 2023. The information gathered from the 2022 surveys is useful in placing the subsequent self-efficacy evaluation in context, because qualitative research can provide a richer understanding of the mindset of the incarcerated students, and open-ended questions allowed them an opportunity to describe their experiences in their own words. Below, we report on the design and considerations common to both studies, then highlight specific differences with regard to participants across the two years. The course curriculum and student:teacher ratio remained virtually unchanged across the 2022 and 2023 programs.}

\subsection{Ethical Considerations}
We followed the ethical guidelines put forth by the Perspectives on Ethical Inquiry~\cite{Perspectives} developed by The Inside-Out Center’s Evaluation and Research Committee\textemdash a leader in evaluations for prison-based educational programming\textemdash to ensure our research exceeded the federal regulations for Respect for Persons, Beneficence, and Justice. Our procedures also ensured that the program instructors were not the researchers, to help mitigate any concerns of coercion or other negative perceptions, and the course instructors were never aware of the students' participation status. These procedures were communicated to all participants before the study was carried out. We received approval of our research design from our institution's Institutional Review Board (IRB). {Incarcerated people are considered a vulnerable population for research purposes, but participation in this survey posed only minimal harm because their participation in the education program was voluntary, participation in the survey was voluntary, the questions did not ask about sensitive topics, and their names were kept confidential. All students were asked to participate if they chose to provide responses, and no responses were excluded from the study. Participants were not compensated for participation in the survey, and the instructors observed secure data handling procedures. } 

\subsection{Facility and Participant Selection}

We carried out our studies with all 5 participating correctional facilities, which we refer to here alphabetically as Facilities A through E. Facility A included participants from both men's and women's units, including residents across low, medium and high security custody categories. Facility B is a medium security facility, and included only women. Facility C is a minimum security facility, and also included only women. Facility D is a medium security facility, and included participants from both men's and women's units. Finally, Facility E is a mixed medium and minimum security facility, and included only men. {Specific student enrollment in the class was controlled by the correctional facilities' education staff, who filter enrollment based on conflicts in students' schedules and recent misconduct; participation in our study was offered to all enrolled students.} The studies were carried out as virtual {surveys} with consenting participants. All participant data is aggregated and anonymized, and we received approval of our research design from our institution's Institutional Review Board (IRB).

{We selected the 5 participating correctional facilities in an effort to reflect the representation in the general prison population. In particular, the participating facilities covered a broad range of security classifications, including a jail, a reentry center, as well as minimum and medium security prisons. This selection was also informed by the values expressed by HCI researchers that "participation of underrepresented minorities must be sought in all of our activities"{~\cite{ogbonnaya2020critical}}. With our original partners exhibiting under-representative populations of Black students (at 2\% and 9.5\%, respectively{~\cite{blackpopulationdata}}), we expanded our partnership in 2022 to reach a more racially representative group of students. Our 2023 study had 29.4\% of respondents identifying as Black, 8.7\% as multiracial, with 8.8\% preferring not to say. This is approximately reflective of the US general prison population, which is 34\% Black according to a 2014 study{~\cite{naacp}}. \textit{Design Justice}{~\cite{Costanza-Chock}}, a practice aimed at ensuring "a more equitable distribution of design’s benefits and burdens" and "meaningful participation in design decisions", also influenced our selection of participating correctional facilities. Because women constitute a relatively small portion of the U.S.~prison population, women are at higher risk of being discounted in correctional research aimed to foster positive post-release outcomes. To reach these students, our program partnered with women's correctional facilities. In our 2022 and 2023 programs, the percentages of initially enrolled students who identified as women were 52.4\% and 49.1\%, respectively. Drawing further from Design Justice practices, our program also included student alumni from previous years who returned to serve as TAs, contributing to the program design and content. } 

\subsection{Participant Demographics}


\subsubsection{2022 Study}
In the 2022 program, there were 45 students enrolled at the beginning of the course. The only demographic data collected in this iteration of the study was gender, with 22 (52.4\%) of the initial 42 respondents identifying as women and 20 (47.6\%) of the initial respondents identifying as men. Of the initially enrolled students, 34 (75.7\%) both remained enrolled at the end of the course and gave consent to complete the post-course survey. 

\subsubsection{2023 Study}
In the 2023 program, 57 students were enrolled at the beginning of the course. {Among the students initially enrolled in the course, 28 (49.1\%) identified as women and 29 (50.9\%) identified as men.} Of the initially enrolled students, 37 (64.9\%) provided consent and completed the initial survey. At the end of the course, 32 (86.5\%) of the participants remained enrolled and consented to complete the final survey. This was an overall participation rate of 56.1\%. Demographic data was also collected at the 2023 initial survey. The average age of the participants was 35.15 years (SD = 7.86) and all of them reported a preference to speak and read in English. Almost half of the participants (48.6\%) had completed high school, while 42.9\% had completed some college, 5.7\% vocational school, and 2.9\% less than a high school diploma. Few participants (11.4\%) identified as being of Spanish, Latino/Latina/Latinx, or Hispanic origin (5.7\% preferred not to say). Lastly, a slight majority of the participants identified as White (52.9\%), while 29.4\% identified as Black, 8.7\% as multiracial, and 8.8\% preferred not to say.

\subsection{Structure and protocol}

\subsubsection{2022 Study}
Our 2022 study consisted of a post-course survey, administered virtually at the end of the program. This survey was carried out with consenting participants, using both dichotomous and open-ended questions.  {Instructors chose these questions to better understand the individual and organizational needs of students, their concerns, interests, prior experience, and their overall assessment of the value of the course, with the intention of using this information to improve the classroom experience for future students. Open-ended questions and Likert-type questions were used to develop global measures of satisfaction with the course. Students were asked a variety of questions about their expectations for the course and their prior experience with programming, but one question asked them about social issues they cared about that they would like to develop solutions to. Students prioritized topics such as poverty, homelessness, abortion laws, the legal system (especially sentencing guidelines), climate change, freedom of speech, their communities, gun control, adult and child abuse, bullying in schools, breast cancer, addiction resources and issues related to social justice/racism. Student responses to open-ended questions were analyzed for potential themes and categorized according to their predominant message (e.g. developing a feeling of empowerment, overcoming initial inhibitions, and perceived limits of the course effectiveness). An important benefit of these responses is giving the students an opportunity to speak for themselves, adding context to our quantitative measures of self-efficacy with their personal experiences, and giving us insights into the course from their unique perspective.} We conducted a thematic analysis of responses and detail our findings in the following section.

\subsubsection{2023 Study}

We measured two forms of self-efficacy, general and computer-specific. General self-efficacy is a widely used measure of a person's self-belief in their competence to tackle novel tasks and to cope with adversity across a range of stressful or challenging encounters, which research suggests is a universal construct linked to optimism, self-regulation, self-esteem and academic performance~\cite{luszczynska2005general}. In our study, we measured self-efficacy using the General Self-Efficacy (GSE) scale~\cite{schwarzer1995generalized}. The use of specific self-efficacy scales for computers and computer programming are now also growing rapidly, partly in response to computer literacy requirements implemented for middle school students in some countries \cite{tsai2019developing}. Previously, HCI researchers have developed new measures of self-efficacy as proxies for skill among secure software developers \cite{votipka2020building}. For our course, we use the computer self-efficacy measure developed by Howard et al.\cite{howard2014creation}, which reports excellent psychometric properties and internal reliability, as well as strong evidence for validity.

\textbf{General Self-Efficacy Survey.} As in Allred et al.~\cite{allred2013self}, we measured self-efficacy using the General Self-Efficacy (GSE) scale~\cite{schwarzer1995generalized}. The GSE scale measures a person's ability to cope with daily hassles and also captures adaptation following major stressful life events. The GSE scale is a 10-item survey where participants respond to how true each item is for them. Originally, the GSE response scale was from 0 (not at all true) to 4 (exactly true). In our survey, we modified the response scale to a 5-point scale from 1 (not true at all) to 5 (completely true) to better match the response scale on other survey items and reduce confusion for the participants. A total score was calculated as the sum of scores for each item, and then normalized for comparison with the computer self-efficacy scores (range: 1-5). Previous studies have reported reliability coefficients (Cronbach’s alphas) from 0.76 to 0.90 and the reliability of this scale in our sample was also high (pre-test $\alpha$ = 0.84, post-test $\alpha$ = 0.81), indicating internal reliability.

\textbf{Computer Programming Self-Efficacy Survey.} As recommended by Schwarzer \& Jerusalem~\cite{schwarzer1995generalized}, we included a measure to tap into behavior change specific to the course content. We used the Computer Programming Self-Efficacy Scale (CPSES)~\cite{tsai2019developing}, a measure based on a framework for distributed computational thinking that encompasses the elements needed to successfully complete a programming task~\cite{berland2011collaborative}. The CPSES comprises 16 items of five sub-scales: logical thinking (the ability to write a program using logical conditions), cooperation (the extent to which students perceive a programming task as cooperative), algorithm (the ability to build an algorithm to solve a problem), control (confidence in using a program editor), and debug (the ability to fix errors). Participants respond to how well each item describes them on a scale from 1 (does not describe me) to 5 (describes me extremely well); again, this response scale was modified for ease of use by the participants (original scale was 6 points). We averaged all the items from each subscale to report subscale scores. The reliability for the subscales from the original research ranged from 0.84 to 0.96. The reliability for the subscales in our sample is listed in Table \ref{table:self-efficacy-subscale}, where item numbers refer to the ordering in which questions from a sub-scale were asked. As reported in Table \ref{table:self-efficacy-subscale}, our measures of reliability found that the subscale for logic ($\alpha$ = 0.87), algorithms ($\alpha$ = 0.84), control ($\alpha$ = 0.91), and debugging ($\alpha$ = 0.91) fell withing the range reported in prior research on computer programming self-efficacy~\cite{tsai2019developing}, and only the subscale for cooperation ($\alpha$ = 0.79) fell outside this range. We interpret these findings to mean that our use of the CPSES was appropriate for this study.

\begin{table}[H]
\caption{Reliability of Computer Programming Self-Efficacy Subscales at Pre- and Post-Test}
\begin{tabular}{p{0.1\textwidth}p{0.1\textwidth}p{0.09\textwidth}p{0.09\textwidth}}
Subscale & Item \newline Numbers & $\alpha$ (Pre-Test) & $\alpha$ (Post-Test) \\ 

\hline
\\[-2mm]
 Logic &  1, 6, 11, 16 & 0.87 & 0.88 \\
Cooperation & 2, 7, 12 & 0.79 & 0.71 \\
Algorithms &  3, 8, 13 & 0.84 & 0.89 \\
Control & 4, 9, 14 & 0.91 & 0.80 \\
Debug & 5, 10, 15 & 0.91 & 0.88
\end{tabular}
\label{table:self-efficacy-subscale}
\end{table}


\section{Results}

{In this section, we first report on the thematic analysis of our 2022 post-course surveys, and then report the results from the statistical analysis from our 2023 study.}

\subsection{Thematic Analysis of Participant Feedback}

This section offers a thematic analysis of the qualitative data gathered from post-course surveys completed by 34 incarcerated students in our 12-week, college-accredited web design course from 2022. We focus on understanding the impact of HCI elements on self-efficacy and learning experiences within the unique carceral context. 

\subsubsection{"I Can Do It": Discovering Self-Potential and Self-Efficacy:} 
\hfill \break
\textbf{Empowerment through Education.}
Participants overwhelmingly reported an increase in self-confidence, attributing their newfound self-belief to the course. One participant wrote, \textit{``it shows me that I can do it. It's never too late to learn something new''} (p4)\footnote{p indicates participant number, ranging from 1 to 34}. Others said, \textit{``This class has given me more confidence in myself as well as using a computer''} (p10), and \textit{"for me the best part was learning a computer all over again after not using one in 21 years and then learning code at the same time."} (p16). These sentiments were echoed by others: \textit{``I never thought I could do anything like this''} (p11), \textit{"My self-confidence is on another level"} (p27), \textit{"it showed me how to do something that I felt was so hard"} (p29) and \textit{``one of the best parts of this class were the 'aha!' moments where it's like 'I can code! I’m really doing it!''} (p15). Another student ascribed further transformative effects to the course, writing \textit{"this class has shown me that I am human again and I deserve to have a better quality of life post-incarceration"} (p32). The class also appeared to motivate students to pursue further education, despite limited previous computing experience: \textit{"As computer illiterate as I was, learning web design has sparked my thirst for attaining knowledge from this day forward."} (p23). This theme underscores the transformative power of computing tools in enabling learners to realize their own capabilities.
\hfill \break
\textbf{Overcoming Initial Inhibitions: Shifting from Shyness to Speaking Up.}
The course also appeared to help individuals overcome their initial hesitations or shyness, contributing to increased self-efficacy. As one participant put it, \textit{``at first, I was a shy person and now I’m not afraid to speak up''} (p16). Another stated that the course \textit{``challenged me to stay focused and learn something new that I had zero background in''} (p20). Yet another mentioned, \textit{``at first I ain't think I could do [it], but now I know anything you put your mind to you can really do it''} (p21). Another student's comment suggests the positive experiences that may arise from sharing a virtual classroom with students of other genders: \textit{“this is the first class I have taken where there was interaction with males. I have been incarcerated for 21 almost 22 years and even the teachers here are 90\% female so my confidence around the male population is a big fat ZERO. With this class it was great to get feedback from the guys about my site, which I was terrified by the thought of the guys giving us feedback at first, but then it was great. It gave me a little more confidence [and] it was definitely not something that I was expecting to do for my first computer experience in 21 years”} (p14). This suggests that while web programming courses like ours are seldom offered in correctional facilities, and incarcerated students typically lack experience in computing, giving those students the opportunity to engage with these new materials can invoke positive changes in their self-belief. 

\subsubsection{Counter Narratives: Limitations to Increased Self-Efficacy}
\hfill \break
\textbf{The Role of Faith and Pre-Existing Confidence.}
A few participants did not report a boost in self-confidence due to the class. One such comment was, \textit{``my confidence [only comes from] the almighty God JESUS MY savior teacher and everything else''} (p5). This highlights how computing programs' impact on self-efficacy might be affected by individual beliefs.
\hfill \break
\textbf{Previous Exposure to Programming.}
Another participant who did not report increased self-confidence had previous exposure to coding, stating, \textit{``I mean... [course number of previously taken programming class] did most of that...''} (p9). This suggests while exposure to computing programs can boost self-confidence, subsequent exposure may have diminishing impacts.

\subsubsection{Perceived Value: Beyond Coding Skills}
\hfill \break
\textbf{Creating Meaningful Websites.}
The students also found great value in the real-world application of their skills, particularly in creating meaningful websites. For instance, \textit{``it was really awesome to see something that I built up from nothing but an idea and some characters turn [out] as well as this did''} (p9). Others echoed this sentiment, one student writing \textit{``the best part of the class was working on my own website and learning all the codes and how to use them. I have a whole new outlook on websites now''} (p10), and another commenting that the best part of the class was \textit{"the opportunity to support victims of domestic violence through the creation of this website."} (p13).
\hfill \break
\textbf{Tailored Instruction and Inclusivity.}
Participants praised the inclusive and personalized nature of the instruction, including the use of breakout rooms for more individualized learning. One student's comment exemplified this theme, offering \textit{"the best part about this class was that I received all the help needed to be able to solve some obstacles I encountered throughout the class."} (p29). Additional class highlights suggested by students were \textit{"working with people who took their time to explain how to code and made sure we understood"} (p20), \textit{"a welcomed learning environment that let me escape for 4 hours a week"} (p14), \textit{``communication with TAs...individual instruction when needed''} (p3), and \textit{``the non-judgement completely inclusive atmosphere''} (p11). One student summarizes the sentiment shared by their peers, writing \textit{"I think the best part of this class was the fact that there were no expectations put on us other than that we did our best. I also really appreciate the one-on-one time we were able to have with our TAs"} (p7). Another student corroborated this sentiment, stating "\textit{the Brave Behind Bars program embodies all of the principles of Social Justice while demonstrating kindness in their promotion of underserved and underrepresented populations in technological fields. [It] was a joy to watch peers of mine who consider themselves technologically illiterate, light up and say “I’m doing it! I made a web site!” Sharing this experience as a global citizen of a synchronous online classroom that integrated cohorts from [three different states] was a wonderfully inclusive collaboration that made web/tech development feel accessible to all}" (p15).

\subsubsection{Summary}
In summary, the thematic analysis of the 2022 class feedback suggests the predominantly positive impact of web programming in boosting self-efficacy among incarcerated students, while also highlighting instances where this was not the case. By weaving in the direct quotes from the participants, we provide a nuanced and rich account of the student experience. While offering critical insights for future computer literacy classes designed for marginalized user groups, this study also motivated a more quantitative analysis of self-efficacy which we incorporated into our 2023 study, discussed next. 

\subsection{Statistical Analysis}

Table \ref{table:survey-stats} shows the sample sizes (n), means (M), and standard deviations (SD) for student outcomes, for both pre-test and post-test. Note that differences in sample size reflect incomplete data. The mean scores measuring general self-efficacy and the CPSES subscales all increased between pre- and post-test, but the relatively large standard deviations indicate substantial variation in scores, perhaps due to our small sample size (n=28-36). 

Table \ref{table:survey-results} shows the results for one-sided paired t-tests to determine whether student outcomes at the end of the term were improved relative to the start of the term. For this test, any participant who did not complete both pre- and post-tests for an item were excluded from the analysis. This left 15 individuals who completed a pre-test and a post-test for measures of general self-efficacy, logic CPSES, and algorithms CPSES, and 14 individuals who completed the CPSES subscales for cooperation, control, and debugging. We applied an alpha level of 0.05 for all t-tests. The mean difference shows in this case that all CPSES subscales increased between pre- and post-test, while the general self-efficacy score fell marginally by 0.01, but exhibiting the largest variability in our data. Contrary to our hypothesis, neither general self-efficacy nor any CPSES subscales were statistically significantly different at post-test compared to pre-test. The difference in means test found that there was no statistically significant difference between the mean scores of self-efficacy and CPSES subscales, except for self-efficacy in algorithms (p value = 0.01). We cannot reject the null hypothesis to claim that the program increased students' self-efficacy, but the small sample size provided limited statistical power to make such a determination. Replication studies can incrementally add to the research on this subject and provide a basis for meta-analyses with increased statistical power.

\begin{table}[t]
\caption{Sample sizes, means, and standard deviations for student outcomes at pre-test and post-test}
\begin{tabular}{lllll}
\multicolumn{1}{c}{Outcome} & \multicolumn{2}{c}{Pre-test} & \multicolumn{2}{c}{Post-test} \\ \hline
\multicolumn{1}{c}{} & \multicolumn{1}{c}{n} & \multicolumn{1}{c}{M (SD)} & \multicolumn{1}{c}{n} & \multicolumn{1}{c}{M (SD)} \\ \cline{2-5} 
General Self-Efficacy & 34 & 4.08 (0.563) & 32 & 4.18 (0.480) \\[2mm]
\multicolumn{5}{l}{Computer Programming Self-Efficacy} \\[1mm]
\qquad Logic &  33 & 3.12 (1.16) & 28 & 3.40 (1.04) \\
\qquad Cooperation & 35 & 3.11 (1.10) & 28 & 3.44 (0.85) \\
\qquad Algorithms &  36 & 2.44 (1.14) & 28 & 2.88 (1.14) \\
\qquad Control & 34 & 3.06 (1.33) & 28 & 3.27 (1.00) \\
\qquad Debug & 34 & 2.73 (1.30) & 28 & 3.10 (1.05) 
\end{tabular}
\label{table:survey-stats}
\end{table}

\begin{table}[t]
\caption{Results of the paired samples test for each outcome variable}
\begin{tabular}{lccc}
\multicolumn{1}{c}{Outcome} & \multicolumn{1}{c}{Mean Difference} & \multicolumn{1}{c}{t-Statistic} & \multicolumn{1}{c}{p-value} \\ \hline
\\[-2mm]
General Self-Efficacy & -0.01 & 0.144 & 0.56 \\[2mm]
\multicolumn{4}{l}{Computer Programming Self-Efficacy} \\[1mm]
\qquad Logic &  0.23 & -1.18 & 0.13 \\
\qquad Cooperation & 0.27 & -2.79 & 0.15 \\
\qquad Algorithms &  0.58 & -2.79 & 0.01 \\
\qquad Control & 0.26 & -1.13 & 0.14 \\
\qquad Debug & 0.48 & -1.51 & 0.08
\end{tabular}
\label{table:survey-results}
\end{table}



Qualitative data from the 2022 course also suggested that the more experience participants have, the less their self-efficacy would improve in subsequent courses. We measured previous experience in two ways: the number of previous programming courses and the number of previous courses that used computer applications. Although we planned to conduct a one-way analysis of variance to determine if changes in self-efficacy may depend on prior experience, we did not have sufficient data to conduct these analyses.



\section{Discussion}

Our thematic analysis of incarcerated students' experiences of our web programming course was overwhelmingly positive. In particular, the themes suggested that students experienced a strongly positive impact in boosting self-efficacy. First, this was suggested in terms of general self-efficacy, including statements such as "\textit{I never thought I could do anything like this}", and "\textit{it showed me how to do something that I felt was so hard}". Second, this impact was also suggested in terms of digital literacy (measured using computer self-efficacy), with statements such as "\textit{this class has given me more confidence in myself as well as using a computer}" and "\textit{one of the best parts of this class were the ’aha!’ moments where it’s like ’I can code! I’m really doing it!}". {The positive responses on students' self-reported motivation revealed by our thematic analysis suggests important design considerations for digital literacy courses in incarcerated environments. While the diverse learning needs of heterogeneous groups of incarcerated students are typically not met by prison education programs{~\cite{czerniawski2016race}}, the low student:staff ratio established in our program (2.7:1) allowed our staff to work with students on an individual basis. We posit that our program design was thereby able to motivate self-efficacy by catering more directly to students' individual needs and paces, accelerating digital skill adoption; a key criterion prescribed by Barros et al.{~\cite{barros2023learning}} for prison contexts in particular. Further, the transformative experiences reported by students under the theme \textit{"I can do it: Discovering Self-Potential and Self-efficacy"} suggest compatibility with findings by Gray et al.{~\cite{gray2019transformative}}, who report that active dialogue facilitated by our student:staff ratios is key to fostering students' sense of self-determination and confidence. Touching on the wider implications of these results, Gray et al. note that while these transformations begin with the individuals, they filter out to their correctional institutions, and reinforce values of acceptance and inclusion that extend to the community and society more widely. 

Another component to the class structure that may have influenced the program's success in stimulating self-efficacy is its heavy focus on a project-based open-ended design problem, which studies demonstrate are overwhelmingly endorsed by students~{\cite{docherty2001innovative}}. Moreover, our students were given the liberty of choosing project topics for their websites themselves, and TAs worked with students to accomplish their self-prescribed goals for its design. Studies have shown that performance accomplishments in the service of one's community leads to increased general self-efficacy{~\cite{ma2021enhancing}}, reflecting the sense of accomplishment that students reported from having created real websites that addressed a social cause important to them individually. Another influencing factor may be the gender parity in our virtual classroom. Prisons in the United States are gender-segregated, chiefly for security reasons. However, our virtual program enabled us to bring men and women into the same virtual room to learn together, and research suggests contact between incarcerated men and women can help fulfill interpersonal needs{~\cite{carcedo2008men}}, increasing satisfaction with the program. Another finding was the apparent uniformity in the positive feedback among participants. Our cohort's diversity in terms of gender, race, and facility security classification suggests that the digital literacy program's effect on self-efficacy may generalize well across prison demographics. The two exceptions noted in the "Counter Narratives" theme were associated with faith and previous exposure to programming however, suggesting that subsequent exposure may have diminishing returns on self-efficacy. Future work should establish how much exposure is necessary to achieve appropriate results. }

To corroborate these qualitative findings from our 2022 study, we undertook a quantitative analysis in 2023 to measure students' general and computer self-efficacy both before and after completing the program, and compared the results. The mean self-efficacy scores increased at the end of the course for both general and computer self-efficacy for our aggregated samples. This increase was also reflected in our paired t-test results for computer self-efficacy. While this was not true of the paired t-tests for general self-efficacy, its decline was by the \textit{least count}, and was the measure exhibiting the highest variability in our study. However, it is important to note that the comparison of measures of self-efficacy before and after the class were not statistically significantly different from one another in our analysis. There are many possible explanations for this, not least of which is volatility due to the small sample size for these initial courses (n=15; n=14 for paired t-tests), where low statistical power reduces the chance that we can detect a true difference in means~\cite{button2013power}. Another concern is that the overall effectiveness of prison-based programs may be overwhelmed by the criminogenic effects of the environment~\cite{austin2017limits}, where external factors may have changed beyond our control. These challenges are congruent with that faced by other literature in carceral environments, where several individual studies report non-significant results~\cite{crippen2009role, davis2009exploration, galyon2012relationship}. These individual studies add valuable data that are typically compiled in meta-analyses~\cite{honicke2016influence} to achieve statistical significance and discuss learnings from the ensemble. Our quantitative study adds to this data, and new courses are currently underway, for which instructors in our program have incorporated feedback from prior students to improve instruction methods and promote student self-confidence.

{By reporting the structure and struggles in our program, we provide a blueprint for how online prison education programs may be operated with positive effect. As reported in Moreira et al.{~\cite{moreira2017higher}} and in Pike and Adams{~\cite{pike2012digital}}, incarcerated students have typically faced a major challenge in the lack of material and human resources required to focus and encourage learning. Researchers have noted that digital skills are a key component to prison rehabilitation in today's digital market{~\cite{zivanai2022digital}}, yet there are today few proven templates for how this can be achieved in practice. Our results suggest that online programs, appropriately staffed with virtual instructors, may offer a meaningful way to connect incarcerated students to digital course content and instructors in a way that circumvents the significant challenges involved in bringing physical materials and staff into prisons. While virtual programs require access to a laptop and a stable internet connection, which remain limited in U.S.~correctional facilities{~\cite{reisdorf2022locked}}, evidence suggests this access is expanding.} A 2011 report from the Institute for Higher Education Policy recommended support for Internet-based delivery of postsecondary education in prisons to address capacity challenges limiting access to post-secondary education~\cite{gorgol2011unlocking}. Similarly, a 2015 report from the RAND Corporation acknowledges that access to technology is increasingly required for carceral educational programs and reentry preparation~\cite{RR-820-NIJ}. To this end, many states have established official policies for regulated computer and internet access~\cite{sd-internet,maine-internet,ohio-internet}, with states like California even providing free tablets to all inmates~\cite{ca-tablets}. In 2022, California announced a roll-out of 30,000 computers with controlled internet access for incarcerated students{~\cite{cdcr}}. With the in-person COVID-19 lockdowns, many facilities previously reticent to enable internet access installed internet infrastructure and relaxed their policies \cite{tabersecond,dunkel2020impact}, paving the way for programs like ours to meet this opportunity. The logistical, security, and administrative challenges that were overcome to deliver our program\textemdash one of the first of its kind in the United States\textemdash are detailed in this manuscript to foster transparency and encourage further research in correctional education. {Future courses will also incorporate formal evaluations of self-efficacy by observing whether their final websites show improvement in their web development skills and a positive outlook on their futures. We also recommend that instructors and researchers develop similar metrics of improvements in self-efficacy in order to contribute to better research methodologies.}

\section{Conclusion}

We have introduced a college-accredited web design program for incarcerated people, designed to foster digital literacy and self-efficacy. Known as \textit{Brave Behind Bars}, this novel 12-week program is deployed virtually and synchronously across 5 correctional facilities in the United States. It brings together men and women from gender-segregated facilities into one virtual classroom to learn and work together, and facilitates highly tailored instruction via low student-staff ratios. Based on literature establishing self-efficacy and digital literacy as key predictors for incarcerated people's success in the modern workplace post-release, we examined the students’ general and computer self-efficacy as a result of participating in the program. To measure self-efficacy qualitatively, we conducted thematic analyses on {surveys} with students that suggested strongly positive increases in self-efficacy, and discuss reasons for this result. To corroborate these findings, we undertook a second study to measure students' general and computer self-efficacy both before and after completing the program. Our quantitative analysis showed that mean self-efficacy scores increased at the end of the course for both general and computer self-efficacy for our aggregated samples, but larger sample sizes are required to achieve statistical significance. The learnings from our qualitative and quantitative analyses will inform upcoming iterations of our program, and future work is planned to provide the data volume required for more comprehensive statistical analyses. Given the formidable challenges in deploying and studying education programs in carceral settings, our work adds important findings that inform the design of correctional programs to increase self-efficacy as a method to combat recidivism.

\bibliographystyle{ACM-Reference-Format}
\bibliography{references}

\appendix









\end{document}